\documentclass[aps,prl,preprint,groupedaddress]{revtex4}
\usepackage{longtable}
\usepackage{graphicx,color}
\usepackage{times}
\usepackage{amssymb,amsmath}
\usepackage{bbm}

\newcommand{\e}{{\rm e}}
\newcommand{\drm}{{\rm d}}
\newcommand{\irm}{{\rm i}}
\newcommand{\beq}{\begin{equation}}
\newcommand{\eeq}{\end{equation}}
\newcommand{\bdm}{\begin{displaymath}}
\newcommand{\edm}{\end{displaymath}}

\begin{document}


\title{Gravity-Gradient Subtraction in 3rd Generation Underground Gravitational-Wave Detectors in Homogeneous Media}

\author{Jan Harms}
\affiliation{University of Minnesota, 116 Church Street SE, Minneapolis, MN 55455, USA}
\author{Riccardo DeSalvo}
\affiliation{California Institute of Technology, East Bridge, Pasadena, California 91125, USA}
\author{Steven Dorsher}
\affiliation{University of Minnesota, 116 Church Street SE, Minneapolis, MN 55455, USA}
\author{Vuk Mandic}
\affiliation{University of Minnesota, 116 Church Street SE, Minneapolis, MN 55455, USA}




\date{\today}

\begin{abstract}
In this paper, we develop a new approach to gravity-gradient noise subtraction for underground gravitational-wave detectors in homogeneous rock. The method is based on spatial harmonic expansions of seismic fields. It is shown that gravity-gradient noise produced by seismic fields from distant sources, stationary or non-stationary, can be calculated from seismic data measured locally at the test mass. Furthermore, the formula is applied to seismic fields from stationary local sources. It is found that gravity gradients from these fields can be subtracted using local seismic measurements. The results are confirmed numerically with a finite-element simulation. A new seismic-array design is proposed that provides the additional information about the seismic field required to ensure applicability of the approach to realistic scenarios even with inhomogeneous rock and non-stationary local sources.
\end{abstract}
 \maketitle

\begin{figure}[ht!]
\centerline{\includegraphics[width=8cm]{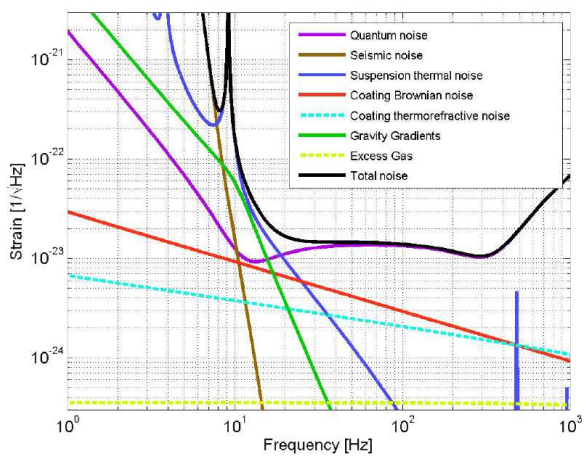}}
\caption{Current estimates for some important noise contributions to Advanced LIGO (generated with the GWINC matlab script).}
\label{fig:adLIGO}
\end{figure}

Seismic waves produce perturbations of the gravity field, which are predicted to cause the so-called Newtonian or gravity-gradient noise (GGN) in gravitational-wave (GW) detectors \cite{Sau1984,BeEA1998,HuTh1998,Cre2008}. Whereas the sensitivity of currently operating detectors is not limited by GGN \cite{LSC2009b,LuEA2006,VIR2008,Tat2008}, second and third-generation detectors will be sensitive to gravity gradients at 10\,Hz and below. Figure \ref{fig:adLIGO} shows the current best estimate for some important noise contributions to the second-generation Advanced LIGO detector.
The GGN curve is based on a model that approximates a typical spectrum of seismic surface waves measured at the Hanford site of the LIGO detectors, and is based on the characteristics of local geology. Third-generation GW detectors will be designed with enhanced sensitivity below 10\,Hz based on improved suspension systems and optimized material properties to mitigate the seismic and thermal noise \cite{RoEA2002,RobEA2004}. Moreover, quantum-non-demolition techniques are being investigated to cancel part of the optical quantum noise \cite{Che2003,KLMTV2001,HaEA2003}. This leaves the question whether the GGN, which is directly imprinted on the test-mass motion without the possibility to build an isolation system, can be decreased. One obvious improvement would be to identify a detector site with a comparatively low level of seismic noise, which also includes the possibility to construct the detector under ground. Underground seismic noise at depths of about 1\,km is known to be an order of magnitude weaker than surface noise above 1\,Hz \cite{Bor2002,CarEA1991,Dou1964}, but further GGN mitigation by two orders of magnitude is required to achieve good sensitivities at frequencies close to 1\,Hz. The solution is to subtract from the GW data an estimate of GGN based on seismic measurements. So far, the assumption was that a 3D array consisting of several hundred seismometers deployed around each test mass extending over several seismic wavelengths would be required \cite{HaEA2009}. In this paper, we will show for homogeneous media that GGN produced by seismic fields from distant sources and from stationary spherical waves can be subtracted using relatively few seismometers positioned at the test mass.

\begin{figure}[t]
\centerline{\includegraphics[width=8cm]{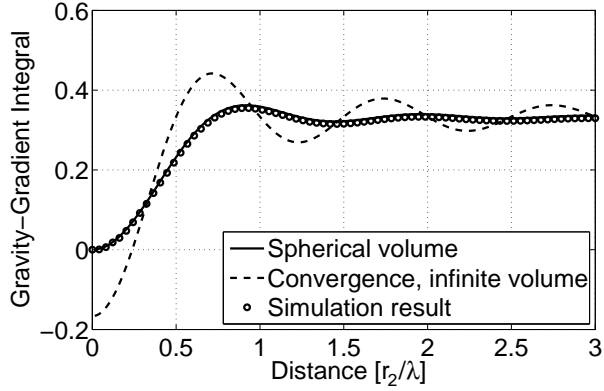}}
\caption{The figure displays a plot of the normalized gravity-gradient integral $s(kr_2)-s(0)$ which appears in Eq.\,(\ref{eq:planeGG}), and the convergence curve of gravity gradients from plane waves in an infinite rock volume as function of integration distance. The two functions differ by the surface term Eq.\,(\ref{eq:Surf}) which needs to be subtracted from Eq.\,(\ref{eq:planeGG}) to obtain the convergence curve. This example demonstrates that in certain problems one needs to take care of surface terms when using the dipole result Eq.~(\ref{eq:planeGG}). Numerical results obtained from a finite-element simulation agree well with the analytic result Eq.\,(\ref{eq:planeGG}). Here, the simulated seismic field was constructed by adding 50 plane P-waves with random directions of propagation. More details can be found in \cite{HaEA2009}.}
\label{fig:ggInt}
\end{figure}

The calculation is based on a plane-wave expansion of the seismic displacement field $\vec\xi(\vec r,t)$. Let us first consider the simple case of a plane pressure wave (P-wave) from a distant source, in which case its frequency $\omega$ and wave number $k\equiv|\vec k|$ are related by $\omega=kc_{\rm p}$, $c_{\rm p}$ being the speed of pressure waves. The P-wave with initial phase $\phi_0$ at the origin $\vec r=\vec 0$ produces longitudinal displacement along the direction $\vec e_k\equiv \vec k/k$:
\beq
\vec\xi^{\,\rm P}(\vec r,t)=\vec e_k\xi_0^{\rm P}\exp(\irm(\phi_0+\omega t-\vec k\cdot\vec r))
\label{eq:planeWave}
\eeq
Assuming that the test mass is located at the origin, the lowest order perturbation of the gravity field can be calculated using the dipole formula \cite{HaEA2009}
\beq
\delta\vec a(t) = G\rho_0\int\drm V\dfrac{1}{r^3}\left(\vec\xi(\vec r,t)-3(\vec e_r\cdot\vec\xi(\vec r,t))\vec e_r\right),
\label{eq:dipole}
\eeq
where $G$ is Newton's constant, and $\rho_0$ denotes the mean density of the rock. The integral is carried out over the entire volume of the medium. Inserting the displacement field of the P-wave, it is straight-forward to solve the integral in spherical coordinates. If one allows for an arbitrary range of integration in radial direction, then the solution is found to be
\begin{eqnarray}
\delta\vec a^{\,\rm P}(t) &= &8\pi G\rho_0\xi_0^{\rm P}\e^{\irm(\phi_0+\omega t)}\vec e_k \left(\dfrac{\cos(x)}{x^2}-\dfrac{\sin(x)}{x^3}\right)\Bigg|_{x=kr_1}^{kr_2}\nonumber \\
&= &8\pi G\rho_0\vec\xi^{\;\rm P}(\vec 0,t)\left(\dfrac{\cos(x)}{x^2}-\dfrac{\sin(x)}{x^3}\right)\Bigg|_{x=kr_1}^{kr_2}\nonumber \\
&\equiv &8\pi G\rho_0\vec\xi^{\;\rm P}(\vec 0,t)\left(s(kr_2)-s(kr_1)\right)
\label{eq:planeGG}
\end{eqnarray}
The inner radius $r_1$ can be interpreted as the radius of a spherical cavity surface. A finite outer radius $r_2$ is needed to evaluate corrections of the result when the integration volume is bounded. The integral as a function of distance is plotted in Fig.~\ref{fig:ggInt}. Instead of considering gravity gradients as a function of the radius $r_2$ of a spherical rock volume, one could ask for the convergence curve of gravity gradients in an infinite rock volume. The difference is that contributions of gravity-gradients from the outer surface at $r=r_2$ need to be subtracted from Eq.\,(\ref{eq:planeGG}). For plane pressure waves, the surface contribution reads
\beq
\begin{split}
\delta\vec a_{\rm surf}^{\,\rm P}(t) = 4&\pi G\rho_0\vec\xi^{\;\rm P}(\vec 0,t)\\
&\cdot\left(\dfrac{\sin(kr_2)}{kr_2}+2\dfrac{\cos(kr_2)}{(kr_2)^2}-2\dfrac{\sin(kr_2)}{(kr_2)^3}\right)
\end{split}
\label{eq:Surf}
\eeq
One does not have to subtract the surface contribution from Eq.\,(\ref{eq:planeGG}) when comparing analytical and numerical results, but it becomes necessary in reality when evaluating gravity-gradient contributions as a function of distance.

The gravity-gradient integral for a plane P-wave, with or without surface term, is of the form $\vec\xi^{\;\rm P}(\vec 0,t)\cdot f(k)$. Therefore, it is possible to add multiple plane P-waves with arbitrary directions of propagation to construct any stationary P-wave field for a specific wave number $k$, and the vector $\vec\xi^{\;\rm P}(\vec 0,t)$ becomes the sum of displacements of all these waves at the origin. So far, we have not made use of the fact that the field Eq.\,(\ref{eq:planeWave}) obeys the plane-wave equation. The gravity-gradient formula Eq.\,(\ref{eq:planeGG}) can be read as an identity in $\vec k$-space, interpreting $k$ as an independent spatial-harmonics variable, and $\vec\xi(\vec 0,t)$, $\delta \vec a(t)$ as amplitudes $\tilde{\vec\xi}(\vec k,t)$, $\tilde{\vec a}(\vec k,t)$ of the respective spatial spectrum. This makes it valid for arbitrary stationary or non-stationary fields from local or distant sources. We will come back to this in the next paragraph. However, the usefulness of the formula for the problem of GGN subtraction is most obvious for fields from distant sources that obey the plane-wave equation enforcing $\omega=kc_{\rm p}$. Translating wave numbers into frequencies, it is clear that the function $f(k=\omega/c_{\rm p})$ can be multiplied in frequency domain after calculating the temporal Fourier transform of Eq.\,(\ref{eq:planeGG}). In other words, it is possible with local data and simple transformations in frequency domain to calculate GGN for a certain range of frequencies. In practice, the applicability is limited to frequencies $f\gtrsim c_{\rm p}/H$, $H$ being the depth of the test mass, since surface effects become important at lower frequencies. A plane-surface correction to Eq.\,(\ref{eq:planeGG}) is required to improve low-frequency predictions of GGN.

A concise form of Eq.~(\ref{eq:planeGG}) can still be maintained if shear waves, which produce GGN exclusively through surface displacement, are added to the total displacement $\vec\xi(\vec r,t)=\vec\xi^{\;\rm P}(\vec r,t)+\vec\xi^{\;\rm S}(\vec r,t)$. For an arbitrary spatial spectrum, one obtains
\beq
\begin{split}
\delta\vec a(t) &= 4\pi G\rho_0\int\dfrac{\drm^3 k}{(2\pi)^3}(s(kr_2)-s(kr_1))\\
&\qquad\qquad\qquad\cdot(3(\vec e_k\cdot\tilde{\vec\xi}(\vec k,t))\vec e_k-\tilde{\vec\xi}(\vec k,t))\\
&= 4\pi G\rho_0\int\dfrac{\drm^3 k}{(2\pi)^3}(s(kr_2)-s(kr_1))\\
&\qquad\qquad\qquad\cdot(2\tilde{\vec\xi}^{\;\rm P}(\vec k,t)-\tilde{\vec\xi}^{\;\rm S}(\vec k,t))
\end{split}
\label{eq:Total}
\eeq
In practice, disentangling the shear and pressure modes is accomplished by a small array of sensors around the test mass, or by instruments which specifically respond to pressure or shear waves (e.g.~strainmeters and rotational instruments, or a measurement of the test-mass position relative to the cavity walls). Once the modes are identified, the integral does not explicitly depend on the direction $\vec e_k$ of propagation. In this way, integration over different directions $\vec e_k$ can be carried out, leaving the integral over the wave number $k$
\beq
\begin{split}
\delta\vec a(t) &= 4 G\rho_0\int\limits_0^\infty\dfrac{\drm k}{2\pi}k^2(s(kr_2)-s(kr_1))\\
&\qquad\qquad\qquad\cdot(2\tilde{\vec\xi}_{\rm d.a.}^{\;\rm P}(k,t)-\tilde{\vec\xi}_{\rm d.a.}^{\;\rm S}(k,t))
\end{split}
\label{eq:totalGG}
\eeq
in terms of the ''direction-averaged'' amplitudes:
\beq
\begin{split}
\tilde{\vec\xi}_{\rm d.a.}^{\;\rm P}(k,t)&=\dfrac{1}{4\pi}\int\drm\Omega\,\tilde{\vec\xi}^{\;\rm P}(\vec k,t),\\
\tilde{\vec\xi}_{\rm d.a.}^{\;\rm S}(k,t)&=\dfrac{1}{4\pi}\int\drm\Omega\,\tilde{\vec\xi}^{\;\rm S}(\vec k,t)
\end{split}
\label{eq:dirAv}
\eeq
where the volume element is written $\drm^3k=k^2\drm k\drm\Omega$. As for the case of a single plane P-wave, Eq.\,(\ref{eq:totalGG}) is easy to evaluate in frequency domain for fields satisfying the plane-wave equation. This result is important, not only because it greatly simplifies the task of subtracting GGN produced by seismic fields from distant sources in homogeneous rock, but also because it serves as starting point to investigate the subtraction problem for fields from local sources in terms of their spatial spectrum. For the discussion of seismic fields from local sources, it will be useful to derive an approximation to the gravity-gradient formula:
\beq
\begin{split}
\delta\vec a (t) &= \dfrac{4}{3} G\rho_0\int\limits_0^\infty\dfrac{\drm k}{2\pi}k^2\left[1-\dfrac{1}{10}(kr_1)^2+\dfrac{\cos(kr_2)}{(kr_2)^2}\right]\\
&\qquad\qquad\qquad\cdot(2\tilde{\vec\xi}_{\rm d.a.}^{\;\rm P}(k,t)-\tilde{\vec\xi}_{\rm d.a.}^{\;\rm S}(k,t)),
\end{split}
\label{eq:approxGG}
\eeq
which neglects terms of order $\mathcal{O}((kr_1)^4)$, $\mathcal{O}((kr_2)^3)$ in the square brackets. The idea is to approximate the cavity as small $kr_1\ll 1$ and the outer surface as large $kr_2\gg 1$. In the limit $r_1\rightarrow 0$ and $r_2\rightarrow\infty$, the formula reduces to
\beq
\delta\vec a(t) = \dfrac{4\pi}{3} G\rho_0(2\vec\xi^{\;\rm P}(\vec 0,t)-\vec\xi^{\;\rm S}(\vec 0,t))
\label{eq:limitGG}
\eeq
The latter equation holds for any field in homogeneous media under the assumption of infinitely small cavity surface and infinitely large outer surface. In other words, whenever the lowest order corrections in Eq.\,(\ref{eq:approxGG}) are negligible, then its limit Eq.\,(\ref{eq:limitGG}) can be applied which is a simple function of local seismic data. Whether the approximation is valid depends on the wave number and the corresponding direction-averaged amplitude. If the spatial spectrum of a field contains non-zero amplitudes $\tilde{\vec\xi}_{\rm d.a.}(k,t)$ for all $k$, then the question is whether the amplitudes are negligible at high wave numbers $kr_1\gtrsim 1$ and small wave numbers $kr_2\lesssim 1$.
\begin{figure}[ht!]
\centerline{\includegraphics[width=8cm]{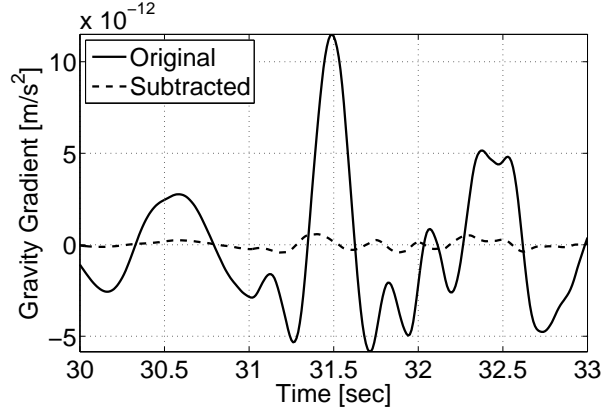}}
\caption{The plot shows part of the simulated GGN from an earthquake measured on February 18th, 2009 at the Kamioka station GIFH10 in Japan. The earthquake had a magnitude of $M=5.2$. Its epicenter was at a distance of 124\,km to GIFH10. Subtracting the GGN estimate Eq.\,(\ref{eq:limitGG}) from a numerical integration of Eq.~(\ref{eq:dipole}), a factor 20 reduction of GGN is achieved.}
\label{fig:earthquake}
\end{figure}

As an extreme example for time-variable distant sources, one can test the applicability of Eq.\,(\ref{eq:limitGG}) with seismic fields from earthquakes. Comparing numerical integrations of Eq.\,(\ref{eq:dipole}) with Eq.\,(\ref{eq:limitGG}) in the case of a real earthquake signal, we found a better than factor 20 reduction throughout the entire waveform. For the East-West displacement and a small stretch of the waveform, the result can be seen in Fig.\,\ref{fig:earthquake}. As one would expect, the error is largest at times when the earthquake signal changes fastest. The subtraction is improved to the level of numerical accuracy, if a frequency-dependent subtraction is applied based on Eq.\,(\ref{eq:totalGG}).

Our next step is to show that Eq.\,(\ref{eq:limitGG}) can also be applied to calculate gravity-gradients generated by seismic fields from stationary local sources. Here one has to take into account that the plane-wave relation $\omega=k c$ does not hold for spherical waves. The wave number $k$ needs to be considered as an independent parameter used in the spatial harmonic expansion of the field. As an example, the stationary P-wave fields from local sources is investigated using the spherical wave originating from $\vec r=\vec r_0$:
\beq
\vec\xi^{\,\rm P}(\vec r,t)=\vec e_r\dfrac{\xi_0}{|\vec r-\vec r_0|}\exp(\irm(\delta_0+\omega t-\vec k_0(\vec r-\vec r_0))
\label{eq:spherical}
\eeq
with $\omega=k_0c_{\rm p}$, $\vec k_0(\vec r-\vec r_0)=k_0 |\vec r-\vec r_0|$, and $\vec e_r\equiv(\vec r-\vec r_0)/|\vec r-\vec r_0|$. Typical local underground sources of seismic waves include Rayleigh-scattering centers, pumping stations, and ventilation systems that produce fields with radiation patterns of varying complexity. Therefore, other types of fields from local sources need to be studied in the future, but here our main intention is to demonstrate how the spatial harmonic expansion can be applied to more general fields. For the chosen spherical wave, one expects a spatial spectrum that extends over a range of wave numbers $k$, with spatial amplitudes peaked at $k=k_0$. A straight-forward calculation leads to the following spatial Fourier transform of Eq.\,(\ref{eq:spherical}), valid for $k\neq k_0$:
\beq
\tilde{\vec \xi}^{\,\rm P}(\vec k,t) = 2\pi\xi_0\e^{\irm(\delta_0+\omega t+\vec k\vec r_0)}\dfrac{\vec k}{k^2}\left[\dfrac{1}{k}\ln\left|\dfrac{k+k_0}{k-k_0}\right|+\dfrac{2k_0}{k^2-k_0^2}\right]
\eeq
The corresponding direction-averaged amplitudes are obtained by inserting the spatial amplitudes into Eq.\,(\ref{eq:dirAv}):
\begin{eqnarray}
\tilde{\vec \xi}_{\rm d.a.}^{\,\rm P}(k,t) &= -2\pi\irm\xi_0\e^{\irm(\delta_0+\omega t)}\dfrac{\vec r_0 }{kr_0}\left(\dfrac{\cos(kr_0)}{kr_0}-\dfrac{\sin(kr_0)}{(kr_0)^2}\right)\nonumber \\ 
&\qquad\qquad\cdot\left[\dfrac{1}{k}\ln\left|\dfrac{k+k_0}{k-k_0}\right|+\dfrac{2k_0}{k^2-k_0^2}\right]
\end{eqnarray}
The spectrum is plotted in Fig.~\ref{fig:Spherical}. The plot shows that the spectrum is indeed peaked at $k=k_0$ and that the amplitudes fall rapidly towards higher and lower wave numbers:
\begin{figure}[t]
\centerline{\includegraphics[width=8cm]{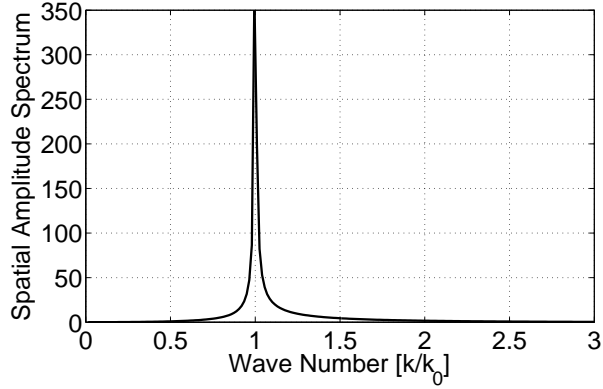}}
\caption{The figure shows a plot of the modulus $|\tilde{\vec \xi}_{\rm d.a.}|$ of the direction-averaged amplitudes of a spherical P-wave. The distance to the center of the spherical wave is $r_0=1$, and the displacement amplitude is normalized to $\xi_0=1$.}
\label{fig:Spherical}
\end{figure}
\beq
\begin{split}
\tilde{\vec \xi}_{\rm d.a.}^{\,\rm P}(k,t)&\quad\stackrel{\stackrel{\scriptstyle k\ll k0}{\phantom{.}}}{\longrightarrow}\quad-\dfrac{8\pi\irm k^2}{9k_0^3}\xi_0\e^{\irm(\delta_0+\omega t)}\vec r_0\\
\tilde{\vec \xi}_{\rm d.a.}^{\,\rm P}(k,t)&\quad\stackrel{\stackrel{\scriptstyle k\gg k0}{\phantom{.}}}{\longrightarrow}\quad-\dfrac{8\pi\irm k_0\cos(kr_0)}{k^4r_0^2}\xi_0\e^{\irm(\delta_0+\omega t)}\vec r_0
\end{split}
\label{eq:Limits}
\eeq
In the following, we test whether Eq.\,(\ref{eq:limitGG}) produces accurate predictions of GGN from spherical waves. As an example, let us assume that the cavity has a radius $r_1=20\,$m and that the test mass is $2\,$km underground, so that Eq.\,(\ref{eq:approxGG}) can be evaluated choosing $r_2=2\,$km. Now, let the wave number of the spherical wave be $k_0=2\pi/(1\,\rm km)$, which corresponds to a 3\,Hz wave propagating in rock with P-wave speed $3\,$km/s. We find that $(k_0r_1)^2\approx 0.016$ and $(k_0r_2)^2\approx 160$. Therefore, at $k=k_0$, both corrections in Eq.\,(\ref{eq:approxGG}) can be neglected given a subtraction goal of two orders of magnitude. Whether amplitudes at higher and smaller wave numbers are negligible, in the sense that Eq.\,(\ref{eq:limitGG}) is still a valid approximation of the exact solution Eq.\,(\ref{eq:totalGG}), is determined by evaluating the integral Eq.\,(\ref{eq:totalGG}) over wave numbers $k$ where the approximation Eq.\,(\ref{eq:approxGG}) does not hold. Alternatively, using a finite-element simulation, one can directly compare gravity gradients from Eq.\,(\ref{eq:dipole}) and Eq.\,(\ref{eq:limitGG}). One finds that the relative deviation depends on the distance of the centers of the spherical waves to the outer surface. The closer the center is to the surface, the greater is the deviation, since low-$k$ surface corrections that are explicitly neglected in Eq.\,(\ref{eq:limitGG}) become more significant relative to the weak gravity gradient produced by more distant sources. The simulation showed that the relative deviation is smaller than 0.02 for all spherical waves that originate at least one wavelength away from the outer surface, and increases to 0.1 for spherical waves which originate very close to the outer surface. This observation may be relevant in practice for strong surface sources. For this reason, a surface array of seismometers is required to guarantee sufficient subtraction of gravity gradients from individual strong surface sources.

In this paper, an approximate time-domain estimate of GGN, Eq.\,(\ref{eq:limitGG}), and a frequency-domain estimate based on Eq.\,(\ref{eq:totalGG}), that is exact for seismic fields from distant sources, have been developed for homogeneous media. If ideal conditions as homogeneity of the rock and stationarity of local sources do not apply, then there are several options to adapt the seismic array and to gather the information needed to improve the gravity-gradient model for realistic seismic fields. In addition to the seismometers at the test mass, sensors can be deployed near large geologic faults and other inhomogeneities that produce significant scattering of the seismic field \cite{Wu1962}. A surface array of seismometers would be necessary to subtract gravity gradients at very low frequencies where the Earth surface cannot be treated as far away from the test mass, and to improve subtraction of strong local surface sources. A dense seismic array around the test mass could be used to measure the spatial amplitudes of the seismic field as a function of time, which would allow us to evaluate Eq.\,(\ref{eq:totalGG}) in a more accurate form \cite{Cap1969}, and improve subtraction of all local sources.

We gratefully acknowledge the support of LIGO, and thank the operators of the Kiban-Kyoshin Network for providing the high-quality earthquake data that was used for this publication.

\raggedright
\bibliography{c:/MyStuff/references}

\end{document}